\documentclass[aps,prc,twocolumn]{revtex4}

\usepackage{graphicx,epsfig}
\usepackage{bm}

\begin{document}
\topmargin .2cm
\title{ELECTROMAGNETIC RADIATION FROM AN EQUILIBRIUM QUARK-GLUON PLASMA SYSTEM}
\author{S.S.Singh$^1$ and  Agam K. Jha$^1$}
\email{sssingh@physics.du.ac.in,akjha@physics.du.ac.in.}
\affiliation{ Department of Physics and Astro-Physics, University of Delhi, Delhi - 110007,INDIA}
\begin{abstract}
  We study the electromagnetic radiation from a hot 
and  slightly strong interacting fireball system of 
quark-gluon plasma using the Boltzmann 
distribution function for the incoming particles and 
Bose-Einstein distribution for gluon in first calculation of electromagnetic
 radiation and
Fermi-Dirac distribution for quark,antiquark and Boltzmann 
distribution for gluon in our second calculation. The thermal photon 
emission rate is 
found that it is infrared divergent for massless quarks 
which are discussed by many authors and regulate this divergence 
using different
cut-off in the quark mass. However we remove 
this divergence  using the same technique of Braaten and Pisarski in the 
thermal mass of  the system by using our model calculation 
in the coupling parameter. 
Thus the production rate of the thermal photon is found to be 
smoothly worked by this cut off technique of our model.
The result is found to be matched with the most of the theoretical
calculations 
and it is in the conformity 
with the experimental results of $200 ~ A~ GeV~  S+Au$ collision of LHC. 
\end {abstract}
\maketitle
 Quantum chromodynamics(QCD), the fundamental theory of strong 
interactions~\cite{wilczek}, predicts a phase transition from confined phase of
hardronic state to deconfined phase of free quark and gluon 
, the so called quark-gluon plasma. The evolution of such deconfined 
state due to the central collisions of two massive nuclei has
been  a subject matter of present day of ultra-relativistic 
heavy-ion collision.
Presumably the early universe was in this state up to about $10\mu s$ 
after the big-bang.Today the core program of ultra-relativistic 
nucleus-nucleus collision is to study the properties of the strongly
interacting matter at very high energy density and temperature so
that it can create this deconfined state of matter in heavy-ion
collision experiments. So the experimental measurement and theoretical 
investigation of electromagnetic observables in heavy-ion collision
constitutes one of the most promising and exciting fields in high 
energy physics. This promising  and exciting observables indicate nothing 
but the emission of high energy photons and leptons~\cite{shuryak} and
these particles have a large mean free path due to the small cross
section for electromagnetic interaction in the plasma. These particles carry
the whole informations about the existence of the plasma and it 
is moreover true that over a large range of expected plasma 
temperature , its radiation can be obsevred throughout its evolution.
So they are considered to be the good signature for the formation of QGP.   
\par
  So far, many authors have studied the production of photon in a QGP 
at finite temperature and this temperature is related to the 
energy density $\epsilon$ given by the stefan boltzmann
$\epsilon~\sim ~ T^{4}$ and the thermodynamic relation for this 
system is given as:
 $T\frac{dp}{dT}-p=\epsilon$; $p=\frac{1}{3}\sigma T^{4}-A T$ 
and the term linear in $T$ gives the non-perturbative effect in 
calculating the pressure and A is the constant value.
Recently T.S.Biro et al.~\cite{biro} derived 
the rate equation describing the chemical 
equilibrium of quarks and gluons
and subsequently by D.Dutta et al.~\cite{dutta} They calculate 
this type of ideal 
fluid dynamics to study the subsequent evolution of kinetically 
equilibrated QGP phase at high energy and temperature. Moreover experimental 
observations~\cite{aggarwal} from 
$Pb-Pb$ collision at $\sqrt{s}=158 ~A~  GeV$at CERN SPS 
indicate favourable interest in this 
electromagnetic probes and the observations show that the 
transverse momentum distribution function of direct photon 
is a significant in the photon induced reaction at the same $\sqrt{s}$
for the transverse momemtum greater than $1.5~ GeV/c$ in the central collision
 but there is doubt  about the QGP formation in the central region at SPS 
energies. This suggests direct photon production as a signal of the QGP phase.
Moreover one important thing is that direct photon production
 is whether dependent on the space time  evolution scenarios of the 
 finite QGP lifetime. For this we focus the photon 
radiation  directly from
 the thermalized quark-gluon plasma at $T=0.25~ GeV$ which 
expands and cools, comes back to hadronic matter with
the production of latent heat which again heats and cools 
and eventuatly at last freezes out into hadrons, mostly as pions.
In this paper,the photon radiation at transition phase around the temperature
$ T=(0.15-0.17)~ GeV $ too,have been calculated and compared with
the hot phase of QGP system i.e the temperature $T=0.25~ GeV$ with the quark
mass depending on the interacting coupling parameter. 
Many investigations  so far have been studied to see hard real 
photon production from the Quark- Gluon Plasma~\cite{kapusta}. 
Based on this results we focus our photon production 
rate at the thermal equilibrium,by using QCD annihilation 
and comptom process between quark, antiquark and gluon by using 
different distribution functions for quark,antiquark and gluon.They 
have been discussed in this paper as different cases for different
 values of temperature $T$ with different coupling parameter used in 
QGP fireball formation. The coupling parameter is calculated with
this parametarised factor $\gamma_{q,g}$$[10]$,
\begin{equation}
  \gamma_{q,g}=\sqrt{2}\sqrt{\frac{1}{\gamma_{q}^{2}}
       +\frac{1}{\gamma_{g}^{2}}},
\end{equation}
where $\gamma_{g}=a \gamma_{q}$ and 'a' is either $6$ or $8$ with
$\gamma_{q}=1/6$.
\par
{\bf Photon production from QGP}: 
The calculation of photon production from a QGP is found to be very
interesting theoretical problem.In the early stage of
universe ,it was a very little understood about photon production just
before the thermalisation process of the system.So we consider 
the thermalised state of QGP after big- bang as the system takes a longer 
time compared to the time sacle associated with the photon production.
Moreover it is stated~\cite{moore}that for the coupling parameter 
$\alpha_{s}<< 1$, 
the result turns out to be slow expansion near the 
equilibrium temperature,even it is not justified condition.
So resulting by this, there are a good number of
research works in photon production from the quark-gluon plasma 
considering the quark, antiquark  annihilation process as well as 
compton process.Most of the time in which photon radiation is used 
by  Boltzmann distribution function by many authors for 
incoming particles
quark and antiquark and for gluon  as Bose-Einstein distribution function.
Thus,now we consider the same approach in which the incoming particles,quark
and antiquark are having the Boltzmann distribution function 
$f_{q}(E_{q})=exp(-E/T)$ and the gluon as Bose-Einstein distribution function 
$f_{g}(E_{g})=\frac{1}{exp(E/T)-1}$ in our first calculation and  
we calculate the photon
emisssion rate produced or photon distribution spectrum 
with QCD coupling parameter~\cite{wong} as follows:
\begin{equation}
E\frac{dN^{ann}}{dP_{\gamma} d^{4}X}=\frac{5 \alpha_{e}\alpha_{s}}{27\pi^{2}} 
T^{2}\exp(-\frac{E}{T})[ln(\frac{4 E T}{m_{q}^{2}})-C_{F}],
\end{equation}
where $C_{F}=C_{Euler}+1+\frac{6}{\pi^{2}}\sum_{n=1}
         \frac{ln(n)}{n^{2}}$
 , $C_{Euler}$  is the Euler number $0.577215$ and $\alpha_{e}=1/137$.
\\
\\
The expression above shows the computed thermal spectrum. Because 
of the quark masses set to be zero,the expression is replaced by 
infrared cut-off $2 k_{c}^{2}$ in the quark mass. The infrared 
divergence is obtained  and  it is regulated by the technique
given by Braaten and Pisarski~\cite{pisarski}. and this is set as 
$k_{c}^{2}=\frac{1}{6} g^{2}T^{2}$ where $'g'$is QCD coupling  
constant and  it is determined from the model where QGP fireball 
is formed~\cite{ramanathan}. The magnitute of $g^{2}$ is given:
\begin{equation}
    g^{2}=\frac{16\pi}{27}
         \frac{1}
          {
           ln(1+\frac{k^{2}}{\Lambda^{2}})
          }
\end{equation}
 with the QCD parameter $\Lambda=0.15~ GeV$ and 
    $ k=(\frac{\gamma_{q,g} N^{\frac{1}{3}} 
      T^{2}\Lambda^{2}}{2})^{\frac{1}{4}}$
 for minimum value of $g$ with $N=\frac{16 \pi}{27}$ 
where $\gamma_{q,g}$ is the phenomological 
flow parameter $[10]$ defined above to take care of the hydrodynamical 
aspects of the hot QGP droplets . It is obviously found that g is 
approximately equal to $1.29 $ which is slightly strong compared to the 
other calculations and $\alpha_{s}=\frac{g^{2}}{4 \pi}$. 
The result produced using this
modification is shown in the Fig.3 for the different values of $T$.
 Again we calculate the comptom process 
i.e $q(\bar{q})g\rightarrow \gamma q(\bar{q})$ using the same 
distribution function for quark and gluon.But the distribution 
function for quark and antiquark are same for Boltzmann 
and Bose-Ein.distribution and the photon production rate 
obtained by this process is given by the following relation.
\begin{equation}
E\frac{dN^{comp}}{dP_{\gamma} d^{4}X}=\frac{5 \alpha_{e}\alpha_{s}}{27\pi^{2}}
T^{2}\exp(-\frac{E}{T})(ln(\frac{4 E T}{m_{q}^{2}})-C_{E}),
\end{equation}
where $C_{E}=C_{Euler}+\frac{1}{2}$.
\\
In a similar  way we plot the figure for this process 
and found to be  similar for both annihilation and 
comptom process for the same distribution function.
In annihilation process, the logarithmic value
 of $ E \frac{dN^{ann}}{dP d^{4}X}$ is slighly higher
than the comptom process.But there is no much difference for higher
 value of $P_{T}$ results.
\par
  In this second calculation we again use the distribution function 
of quark as Fermi-Dirac $f_{q}(E_{q})=\frac{1}{exp(E/T)+1}$and 
Boltzmann distribution for gluon as $f_{g}(E)=exp(-E/T)$.
As we did in the first case,we do the same production rate for annihilation  as 
well as compton process for the different of 
 value of $T$ with QCD coupling parameter.
The production rate for annihilation process is given as:
\begin{equation}
E\frac{dN^{ann}}{dP_{\gamma} d^{4}X}=\frac{10 \alpha_{e}\alpha_{s}}{9\pi^{4}}
T^{2}\exp(-\frac{E}{T})(ln(\frac{4 E T}{m_{q}^{2}})-C_{F1}),
\end{equation}
where $C_{F1}=C_{Euler}+1$
\\
and the comptom process as:    
\begin{equation}
E\frac{dN^{comp}}{dP_{\gamma} d^{4}X}=\frac{5 \alpha_{e}\alpha_{s}}{27\pi^{2}}
T^{2}\exp(-\frac{E}{T})(ln(\frac{4 E T}{m_{q}^{2}})-C_{E}),
\end{equation}
where $C_{E}=C_{Euler}+\frac{1}{2}+\frac{6}{\pi^{2}}\sum_{n=1}\frac{ln(n)}{n^{2}}$.
In both process of annihilation as well as comptom, 
the distribution function used for quark and antiquark 
are same as the system is in thermal equilibrium.The results for this 
annihilation and comptom process are again shown in the Fig. 3 and 4. 
\begin{figure}
\resizebox*{3.1in}{3.1in}{\includegraphics{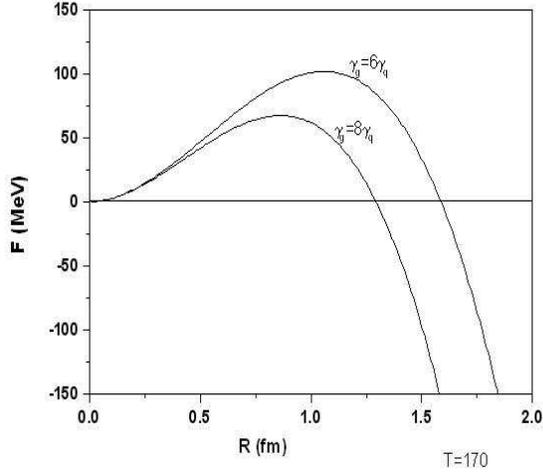}}
\vspace*{0.5cm}
\caption[]{
The free energy $ F~ MeV$ of QGP,  at
thermal temperature $ T=0.17~ GeV$  at different
$\gamma_{q,g}$ with bubble size $R$ in Fermi.
}
\label{scaling}
\end{figure}

\begin{figure}
\resizebox*{3.1in}{3.1in}{\includegraphics{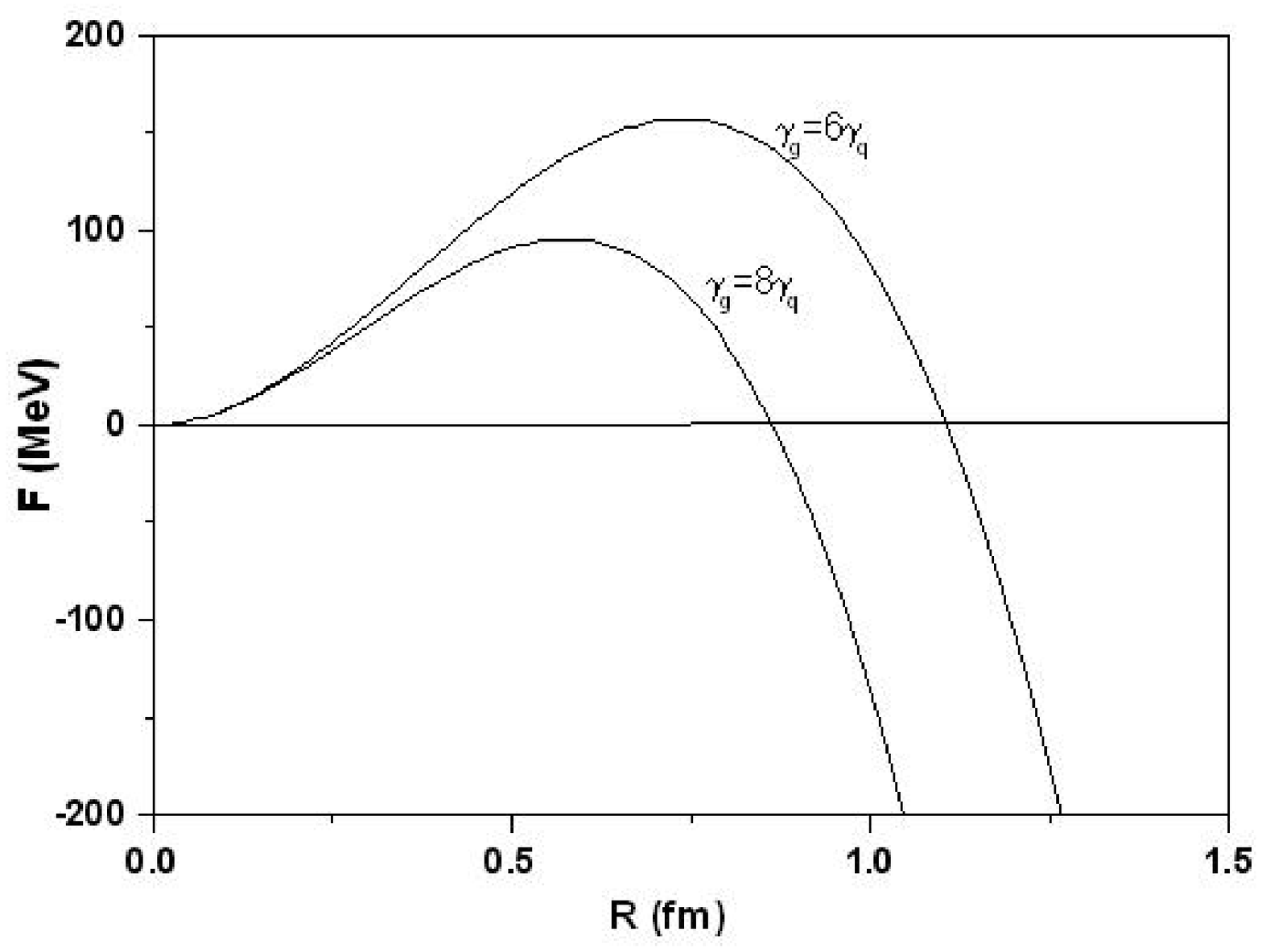}}
\vspace*{0.5cm}
\caption[]{
The free energy $ F~ (MeV)$ of QGP,  at
thermal temperature $ T=0.25 ~GeV$  at different
$\gamma_{q,g}$ with bubble size $R$ in Fermi.
}
\label{scaling}
\end{figure}
                                                                                
\begin{figure}[h]
\resizebox*{3.1in}{3.1in}{\rotatebox{270}{\includegraphics{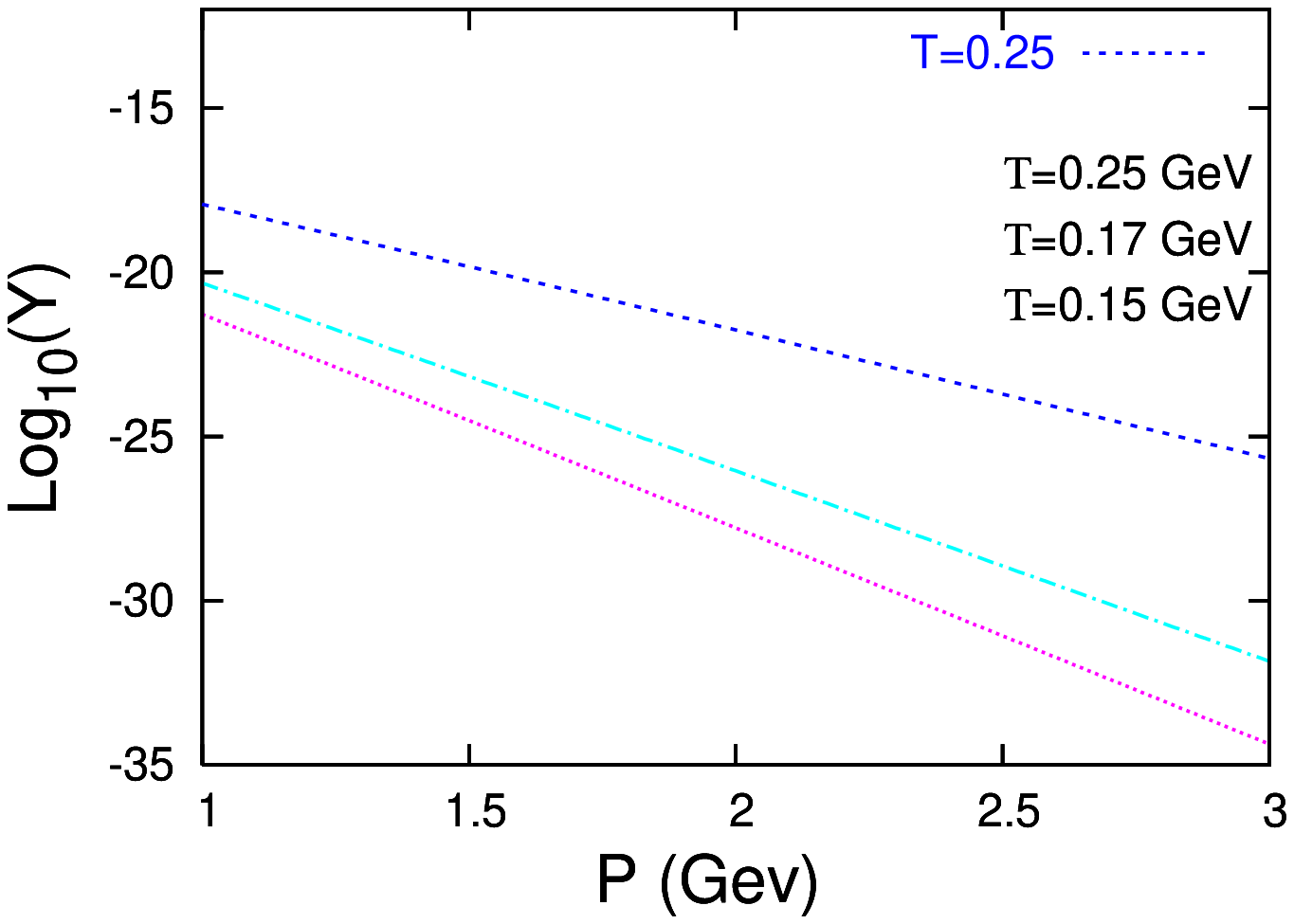}}}
\caption[]{
The photon emission rate through ann,$Y=E\frac{dN^{ann}}{dPd^{4}X}$,
log($Y$) at thermal temperature
$ T=0.25~ GeV,  0.17~GeV ~ and ~ 0.15~ GeV$ at different
$\gamma_{q,g}$ through Boltz dist. for $q $ and $\bar{q}$.
}
\label{scaling}
\end{figure}
\begin{figure}[h]
\resizebox*{3.1in}{3.1in}{\rotatebox{270}{\includegraphics{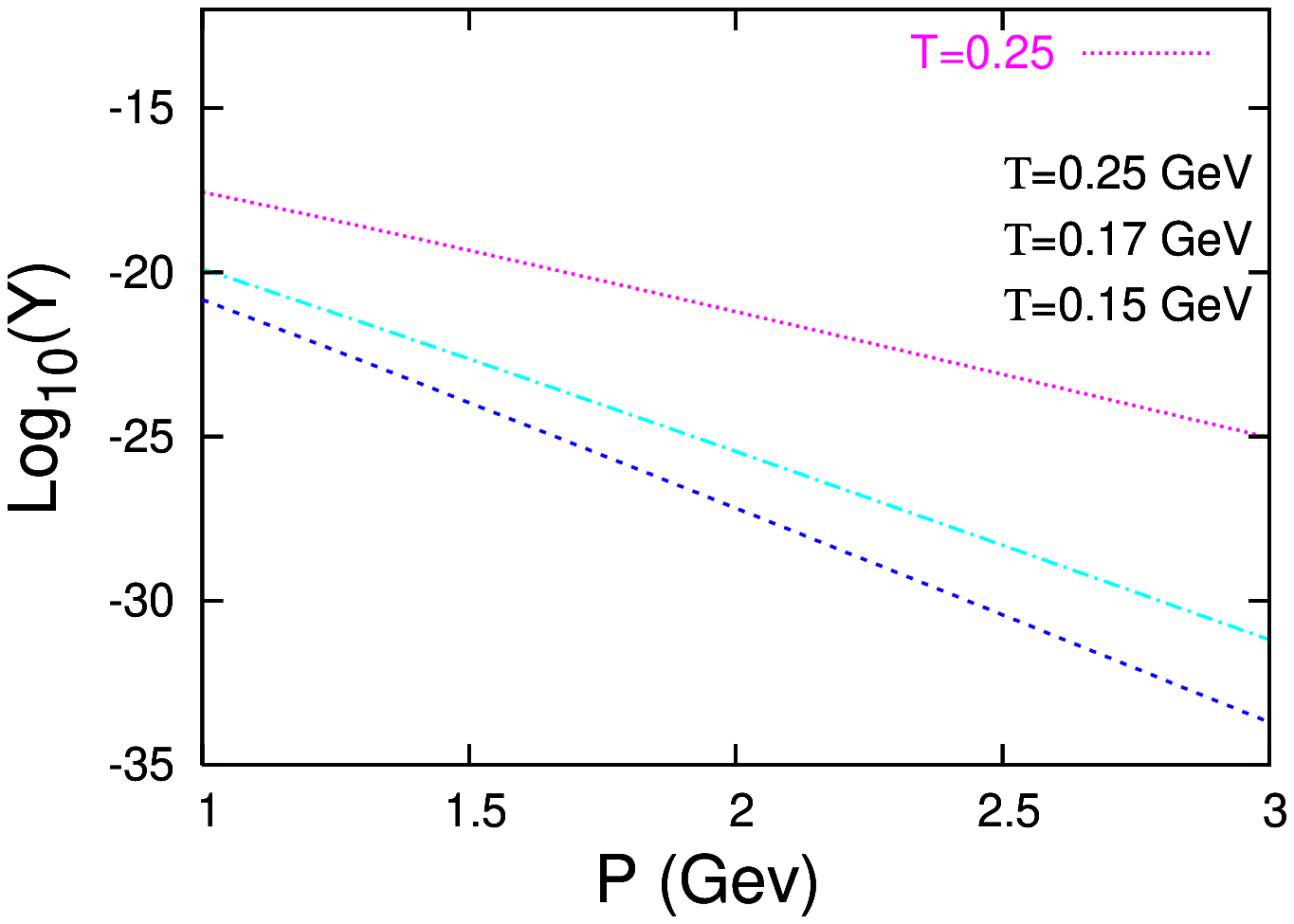}}}
\caption[]{
The photon emission rate through comp,$Y=E\frac{dN^{comp}}{dPd^{4}X}$, log($Y$) at 
thermal temperature $ T=0.25~ GeV,  0.17~ GeV~  and ~  0.15~ GeV$  at different 
$\gamma_{q,g}$ through Boltz.dist.for $q$ and $\bar{q}$.
}
\label{scaling}
\end{figure}
\begin{figure}[h]
\resizebox*{3.1in}{3.1in}{\rotatebox{270}{\includegraphics{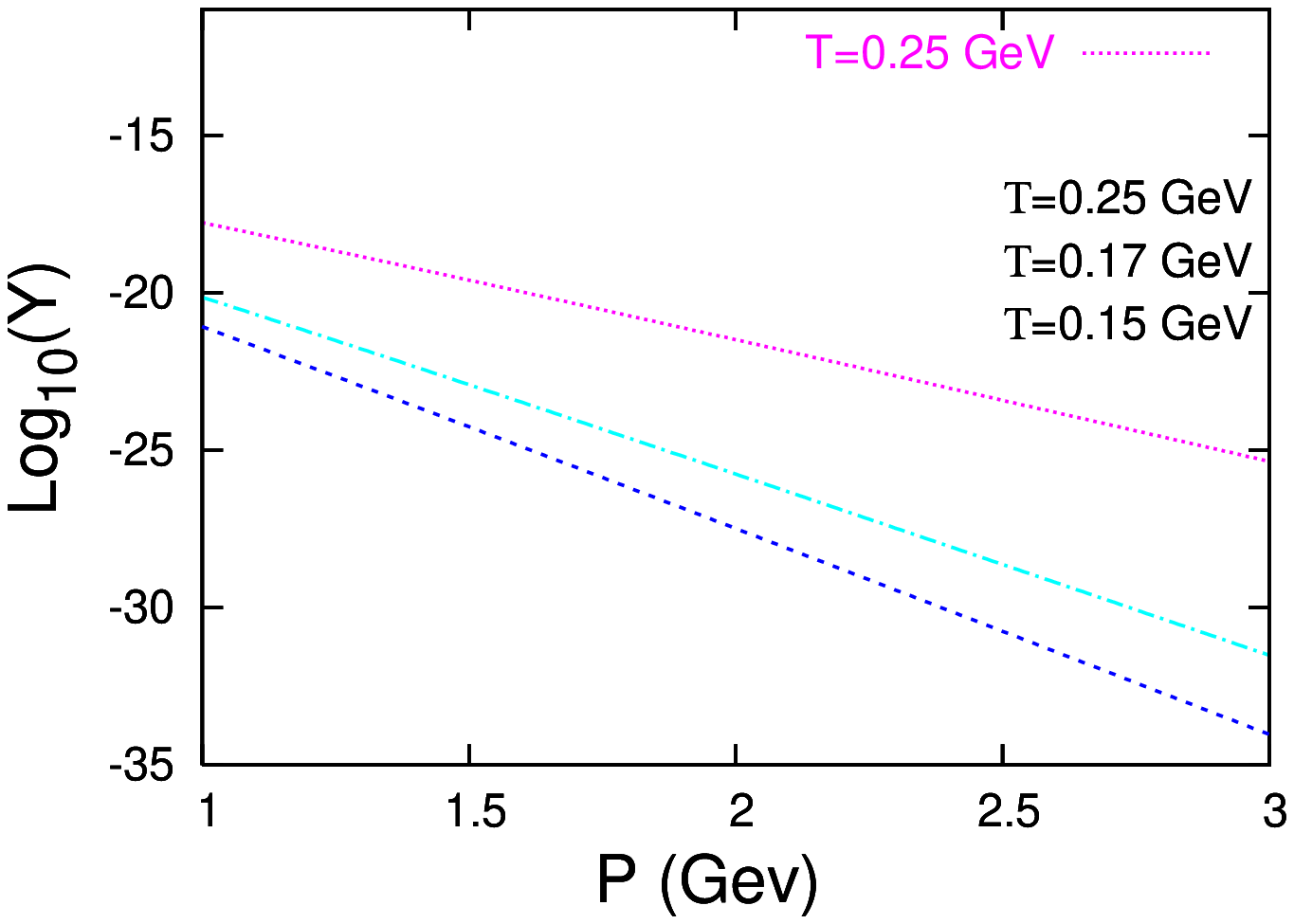}}}
\caption[]{
The photon emission rate through ann, $Y=E\frac{dN^{ann}}{dPd^{4}X}$, 
log($Y$) at 
thermal temperature $ T=0.25~ GeV,  0.17~ GeV ~ and~  0.15~ GeV$  at different 
$\gamma_{q,g}$ through Fermi-dirac for $q$and $\bar{q}$.
}
\label{scaling}
\end{figure}
\begin{figure}[h]
\resizebox*{3in}{3in}{\rotatebox{270}{\includegraphics{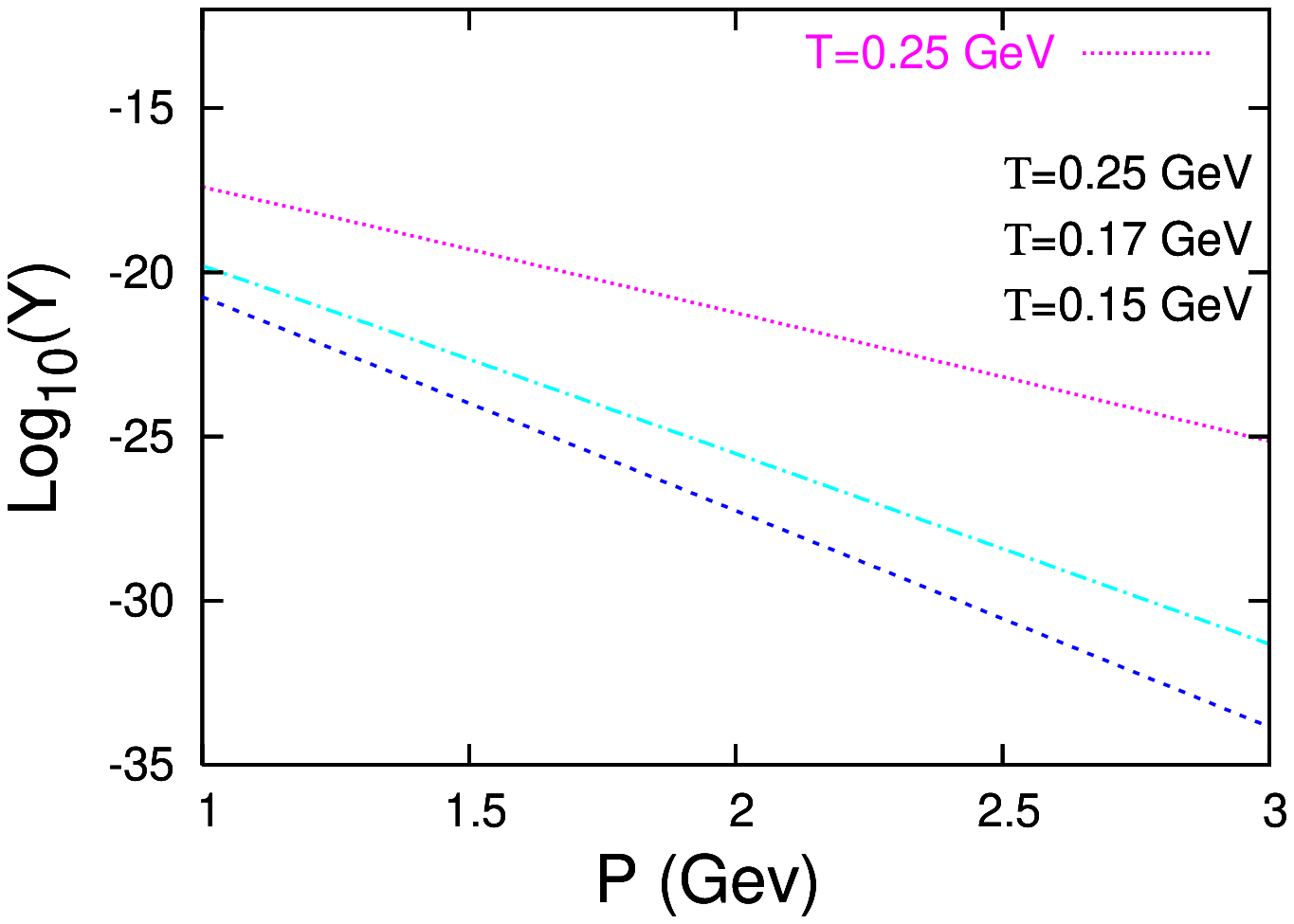}}}
\caption[]{
The photon emission rate through comp, $Y=E\frac{dN^{comp}}{dPd^{4}X}$, log($Y$) at 
thermal temperature $ T=0.25~ GeV, 0.17~ GeV ~ and ~ 0.15~ GeV$  at different
$\gamma_{q,g}$ through Fermi-dirac for $q$ and $\bar{q}$.
}
\label{scaling}
\end{figure}
\begin{figure}[h]
\resizebox*{3in}{3in}{\rotatebox{360}{\includegraphics{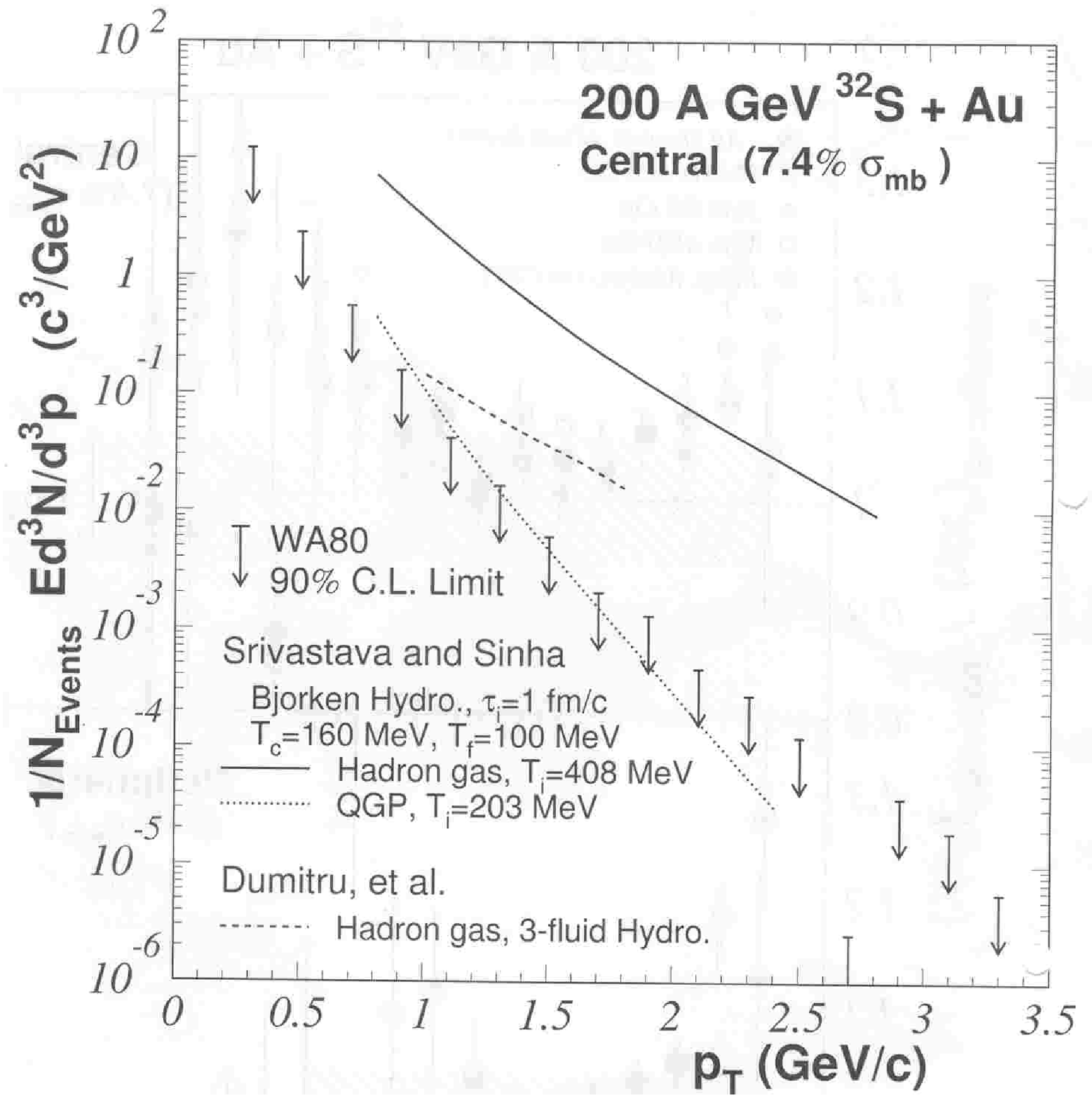}}}
\caption[]{
WA80 collaboration of photon  emission rate at LHC (CERN) 
}
\label{scaling}
\end{figure}

\par
{\bf Results and Conclusions}: In this present paper , we attempt to 
evaluate the photon production rate through two processes using 
the Boltzmann distribution function and Fermi-Dirac distribution 
function for the incoming particles quark and antiquark for hot QGP 
system as well as just around transition phase region for different 
coupling parameters based on $\gamma_{g}=6 \gamma_{q}$ and 
$\gamma_{g}=8 \gamma_{q}$. The photon production rate for both of 
these parameters
 are found to be almost similar in both the distribution functions. 
So the effect of these
 parameters contribute less in finding the photon distribution spectrum.
But the free energy of the QGP at temperature $T=0.17~ GeV$ is found 
to be dependent on this different 
value of coupling constants for both the cases. 
The difference $\gamma_{g}'s$ on $a \gamma_{q}$ is shown 
in the Fig. $1$ with 
the bubble size
 of the QGP fireball. The free energy for $\gamma_{g}=8 \gamma_{q}$
is slighly less for  increase in bubble size compared
with the $\gamma_{g}=6 \gamma_{q}$ , but for the smaller bubble,
free energy 
is  the same. At temperature $T=0.25 GeV$, the result of the free
energy is still higher for $\gamma_{g}=6 \gamma_{q}$
 than the value of $\gamma_{g}=8 \gamma_{q}$. The difference
 between the free energy is more distinct
with increase in bubble size. Moreover, the calculation 
of photon production
 for hot and thermal region of QGP~\cite{cleymans} such as 
the temperature at $T=0.25~GeV$
 which is very hot QGP and round the transition 
temperature at $T=0.17~ Gev $  and  $ 0.15~ GeV $
 show the rate of production different. 
The photon spectrum is found  to be 
 higher compared to those calculated in transition temperature.
Besides this, the production rate spectra is compared with the experiment
results of $200~ A~ GeV~  s+Au $ given by WA80 collaboration at LHC. 
Almost the results are to be quite 
obeying with experimental data with a slight increase in photon spectra.  
Moreover, for higher value of the $P_{T}$ the dissociation rate is slighly
 different with the lower value of the transverse $P_{T}$.Thus,the
photon distribution is found to be very much significant when the
system has higher temperarure. So,photon radiation is considered to be good 
signature for the study of QGP formation.
\acknowledgements
  We are very thankful to Dr. R. Ramanathan and Dr.K.K.Gupta 
for their constructive 
suggestions and discussions.

\end{document}